\documentclass[article]{IEEEtran}
\ifCLASSINFOpdf
\else
\fi
\usepackage[cmex10]{amsmath}
\usepackage{amssymb,amsmath}
\usepackage{algpseudocode}

\usepackage{graphicx}
\usepackage{balance}

\newcommand{\tr}{\mathrm{tr}}
\newcommand{\I}{{\bf \mathrm{I}}}
\newcommand{\numin}{\nu_{\mathrm{min}}}

\begin{document}
\title{Interference Alignment via Message-Passing}

\author{\IEEEauthorblockN{Maxime Guillaud, Mohsen Rezaee, Gerald Matz\\}
\IEEEauthorblockA{
Institute of Telecommunications, Vienna University of Technology \\
Gu\ss hausstra\ss e 25 / E389, Vienna, Austria \\
e-mail: {{\small \texttt{\{guillaud,mohsen.rezaee,gerald.matz\}@tuwien.ac.at}}}} 
%\end{minipage}
%\begin{minipage}{5cm}
%\end{minipage}
}


%Telephone: (800) 555--1212, Fax: (888) 555--1212}


\maketitle

\begin{abstract}
We introduce an iterative solution to the problem of interference alignment (IA) over MIMO channels based on a message-passing formulation. We propose a parameterization of the messages that enables the computation of IA precoders by a min-sum algorithm over continuous variable spaces -- under this parameterization, suitable approximations of the messages can be computed in closed-form. We show that the iterative leakage minimization algorithm of Cadambe et al. is a special case of our message-passing algorithm, obtained for a particular schedule. Finally, we show that the proposed algorithm compares favorably to iterative leakage minimization in terms of convergence speed, and discuss a distributed implementation.
\end{abstract}

%\begin{IEEEkeywords}none.
%\end{IEEEkeywords}


% \fi
\IEEEpeerreviewmaketitle

\section{Introduction}

Interference alignment (IA), introduced in \cite{Gou_Jafar_DoF_MIMO_Kuser_IC_IT2010} for the MIMO interference channel, has received a lot of attention in recent years since it is a key ingredient in achieving the full degrees-of-freedom of the channel. Despite its deceptively simple mathematical formulation, no general closed-form solution to the IA equations is available (although it exists for certain dimensions, see e.g. \cite{Tresch_Guillaud_Riegler_SSP09}). %An iterative algorithm was introduced in \cite{Gomadamit08}, which provides generally satisfactory results. 
In large networks, or when the network topology can not be assumed to be known, distributed implementations are desirable. An iterative distributed implementation relying on over-the-air estimation of interference covariance was proposed in \cite{Gomadamit08}. \\

In this paper, we explore other means of distributed computations of the IA solution, using message-passing (MP).  Message-passing algorithms, and in particular the sum-product algorithm \cite{Kschischang_sum_product_algorithm}, have been used to solve decoding problems in communications. %Since the advent of turbo-codes, they have been an essential tool in algorithm design.
In a nutshell, they provide an efficient way to compute functions involving a large number of variables when the function in question can be decomposed (factorized) into terms that involve only a subset of the variables. %It fact, a number of important algorithms used in communications, such as the Viterbi algorithm, or the Bahl-Cocke-Jelinek-Raviv (BCJR) algorithm, have been shown to be particular cases of message-passing algorithms \cite{Aji_McEliece_generalized_distributive_law_it2000}.\\
MP solutions can be formulated for a variety of problems requiring computations on a commutative semiring, i.e. they are not restricted to the computation of probability distributions \cite{Aji_McEliece_generalized_distributive_law_it2000}. In fact, depending on the choice of the semiring (and the associated binary operations, such as sum, product or minimum), MP solutions are available for Bayesian inference (yielding the belief-propagation approach of ML decoding \cite{Walsh_Regalia_connecting_BP_with_ML}) as well as optimization problems (e.g. the min-sum algorithm \cite{Yedidia_message_passing_inference_and_optimization_2011}).\\

A major motivation for considering MP is that the resulting algorithms are well suited for distributed implementation. This is because for most practical applications, each factor in the function only depends on \emph{local} data (in our application, channel state information). MP has already been considered for beamformer optimization in \cite{Sohn_Lee_Andrews_BP_distributed_DL_beamforming_TWC2011}, where the sum-product algorithm is applied to a function broadly related to the sum-rate. In 
\cite{GHC_distributed_approach_to_precoder_optimization_JASP2013}, the min-sum algorithm is applied to the sum-rate function. The proposed algorithm is restricted to discrete precoders.\\

In this work, we introduce a message-passing technique which optimizes the leakage function associated with interference alignment.  We deal with continuous variables, i.e. we let the precoders and receive filters take any value in the Stiefel manifold, as opposed to the results of \cite{Sohn_Lee_Andrews_BP_distributed_DL_beamforming_TWC2011,GHC_distributed_approach_to_precoder_optimization_JASP2013} where the precoders are chosen from a finite set.
This is motivated by the fact that the number of IA solutions is in general finite \cite{Schmidt_Utschick_Honig_Globecom2010} -- therefore, solving the leakage minimization problem on a discrete subset of the Stiefel manifold will in general not yield an exact IA solution, while solving it on the (continuous) Stiefel manifold will.\\

The contributions of this article are as follows: 
\begin{itemize}
\item We introduce a min-sum algorithm capable of computing the IA precoders in a distributed manner, by associating a suitably chosen graph to the IA problem (Sections~\ref{section_probstatement} and \ref{section_MP}).
\item We propose a parameterization of the messages that enables the use of this algorithm over continuous variable spaces -- under this parameterization, suitable approximations of the messages can be computed in closed-form (Section~\ref{section_continuouscase}).
\item We show that the iterative leakage minimization algorithm of Cadambe et al. is a special case of our message-passing algorithm, obtained for a particular schedule (Section~\ref{section_ILM}).
\item We evaluate numerically the performance of the proposed method, and compare it to the classical iterative leakage minimization (Section~\ref{section_simul}). 
\item We discuss a distributed implementation (Section~\ref{section_distributed}).
\end{itemize}

\subsection*{Notations}
We let $V_{n,p}$ denote the complex Stiefel manifold, i.e. the set of $n\times p$ matrices with complex coefficients and orthonormal columns. $\sim\!\!i$ denotes the set $\{1,\ldots, i-1, i+1,\ldots K\}$ where $K$ is the number of users. For a node $a$ in a graph, $\mathcal{N}(a)$ denotes the set of its neighbors. $\numin \left(\cdot\right)$ returns a truncated unitary matrix spanning the space associated with the $d$ (defined below) weakest eigenvalues of the argument matrix.

\section{Problem statement}
\label{section_probstatement}

Consider the IA equations for the $K$-user interference channel:
\begin{equation}
{\bf U}_i^H {\bf H}_{ij} {\bf V}_j = {\bf 0} \quad \forall i\neq j \in \{1,\ldots,K\}, \label{IAcondition}
\end{equation}
where the ${\bf H}_{ij}$ are arbitrary $N\times M$ complex channel matrices and ${\bf U}_i \in V_{N,d}$, ${\bf V}_i \in V_{M,d}$. Our aim is to solve for the ${\bf U}_i$ and ${\bf V}_j$, given the channel matrices.
Here we focus on the cases where the dimensions are such that IA is feasible for almost all channels with coefficients drawn from continuous distributions 
\cite{Yetis_Jafar_feasibility_conditions_IA_Globecom09}.

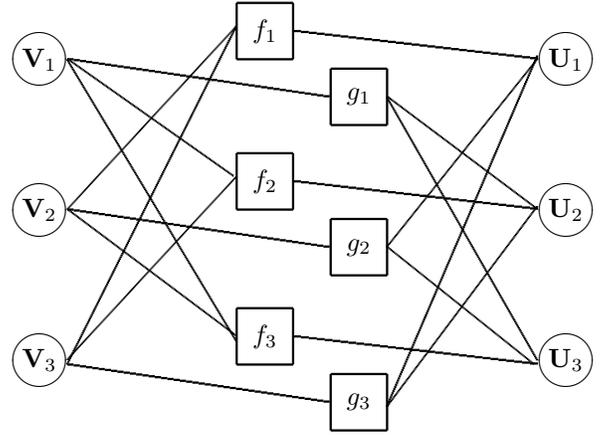
\begin{figure}
\centering \ifx\JPicScale\undefined\def\JPicScale{1}\fi
\unitlength \JPicScale mm
\begin{picture}(85.62,57.5)(0,0)
\linethickness{0.3mm}
\put(38.12,57.5){\line(1,0){7.5}}
\put(38.12,50){\line(0,1){7.5}}
\put(45.62,50){\line(0,1){7.5}}
\put(38.12,50){\line(1,0){7.5}}
\linethickness{0.3mm}
\put(81.88,50){\circle{7.5}}

\put(81.88,50){\makebox(0,0)[cc]{${\bf U}_1$}}

\put(41.88,53.75){\makebox(0,0)[cc]{$f_1$}}

\linethickness{0.3mm}
\put(50.62,48.75){\line(1,0){7.5}}
\put(50.62,41.25){\line(0,1){7.5}}
\put(58.12,41.25){\line(0,1){7.5}}
\put(50.62,41.25){\line(1,0){7.5}}
\put(54.38,45){\makebox(0,0)[cc]{$g_1$}}

\linethickness{0.3mm}
\put(11.88,50){\circle{7.5}}

\put(11.88,50){\makebox(0,0)[cc]{${\bf V}_1$}}

\linethickness{0.3mm}
\put(81.88,30){\circle{7.5}}

\put(81.88,30){\makebox(0,0)[cc]{${\bf U}_2$}}

\linethickness{0.3mm}
\put(11.88,30){\circle{7.5}}

\put(11.88,30){\makebox(0,0)[cc]{${\bf V}_2$}}

\linethickness{0.3mm}
\put(81.88,10){\circle{7.5}}

\put(81.88,10){\makebox(0,0)[cc]{${\bf U}_3$}}

\linethickness{0.3mm}
\put(11.88,10){\circle{7.5}}

\put(11.88,10){\makebox(0,0)[cc]{${\bf V}_3$}}

\linethickness{0.3mm}
\multiput(45.62,53.75)(1.05,-0.12){31}{\line(1,0){1.05}}
\linethickness{0.3mm}
\multiput(15.62,30)(0.12,0.13){188}{\line(0,1){0.13}}
\linethickness{0.3mm}
\multiput(15.62,9.38)(0.12,0.24){188}{\line(0,1){0.24}}
\linethickness{0.3mm}
\multiput(15.62,50)(0.83,-0.12){42}{\line(1,0){0.83}}
\linethickness{0.3mm}
\multiput(58.12,45)(0.16,-0.12){125}{\line(1,0){0.16}}
\linethickness{0.3mm}
\multiput(58.12,45)(0.12,-0.21){167}{\line(0,-1){0.21}}
\linethickness{0.3mm}
\put(38.12,37.5){\line(1,0){7.5}}
\put(38.12,30){\line(0,1){7.5}}
\put(45.62,30){\line(0,1){7.5}}
\put(38.12,30){\line(1,0){7.5}}
\put(41.88,33.75){\makebox(0,0)[cc]{$f_2$}}

\linethickness{0.3mm}
\put(50.62,28.75){\line(1,0){7.5}}
\put(50.62,21.25){\line(0,1){7.5}}
\put(58.12,21.25){\line(0,1){7.5}}
\put(50.62,21.25){\line(1,0){7.5}}
\put(54.38,25){\makebox(0,0)[cc]{$g_2$}}

\linethickness{0.3mm}
\multiput(45.62,33.75)(1.05,-0.12){31}{\line(1,0){1.05}}
\linethickness{0.3mm}
\multiput(15.62,30)(0.83,-0.12){42}{\line(1,0){0.83}}
\linethickness{0.3mm}
\put(38.12,16.88){\line(1,0){7.5}}
\put(38.12,9.38){\line(0,1){7.5}}
\put(45.62,9.38){\line(0,1){7.5}}
\put(38.12,9.38){\line(1,0){7.5}}
\put(41.88,13.12){\makebox(0,0)[cc]{$f_3$}}

\linethickness{0.3mm}
\put(50.62,8.12){\line(1,0){7.5}}
\put(50.62,0.62){\line(0,1){7.5}}
\put(58.12,0.62){\line(0,1){7.5}}
\put(50.62,0.62){\line(1,0){7.5}}
\put(54.38,4.38){\makebox(0,0)[cc]{$g_3$}}

\linethickness{0.3mm}
\multiput(45.62,13.12)(1.05,-0.12){31}{\line(1,0){1.05}}
\linethickness{0.3mm}
\multiput(15.62,9.38)(0.83,-0.12){42}{\line(1,0){0.83}}
\linethickness{0.3mm}
\multiput(15.62,50)(0.17,-0.12){130}{\line(1,0){0.17}}
\linethickness{0.3mm}
\multiput(15.62,10)(0.12,0.13){188}{\line(0,1){0.13}}
\linethickness{0.3mm}
\multiput(15.62,50)(0.12,-0.2){188}{\line(0,-1){0.2}}
\linethickness{0.3mm}
\multiput(15.62,30)(0.16,-0.12){141}{\line(1,0){0.16}}
\linethickness{0.3mm}
\multiput(58.12,3.75)(0.12,0.16){167}{\line(0,1){0.16}}
\linethickness{0.3mm}
\multiput(58.12,3.75)(0.12,0.28){167}{\line(0,1){0.28}}
\linethickness{0.3mm}
\multiput(58.12,25)(0.15,-0.12){130}{\line(1,0){0.15}}
\linethickness{0.3mm}
\multiput(58.12,25)(0.12,0.15){167}{\line(0,1){0.15}}
\end{picture}\\
\caption{Graph obtained for IA over the 3-user, fully connected interference channel.}
\label{fig_3user_graph}
\end{figure}

We reformulate the above IA conditions using the interference leakage metric:
\begin{eqnarray}
\eqref{IAcondition} & \overset{(a)}{\Leftrightarrow}&
\sum_{i=1}^K \sum_{j\neq i} \tr\left( {\bf U}_i^H {\bf H}_{ij} {\bf V}_j {\bf V}_j^H {\bf H}_{ij} {\bf U}_i\right) = 0\\
&\overset{(b)}{\Leftrightarrow} &\sum_{i=1}^K  \underbrace{\sum_{j\neq i} \tr\left( {\bf U}_i^H {\bf H}_{ij} {\bf V}_j {\bf V}_j^H {\bf H}_{ij} {\bf U}_i\right)}_{f_i({\bf U}_i,{\bf V}_{\sim  i})} \nonumber \\
&&+ \sum_{j=1}^K \underbrace{\sum_{i\neq j} \tr\left( {\bf V}_j^H {\bf H}_{ij}^H {\bf U}_i {\bf U}_i^H {\bf H}_{ij}^H {\bf V}_j\right)}_{g_j({\bf V}_j,{\bf U}_{\sim j})} =0.
%&\overset{(c)}{\Leftrightarrow} &\sum_{i=1}^K \underbrace{ \tr\left( {\bf U}_i^H \left[ \sum_{j\neq i} {\bf H}_{ij} {\bf V}_j {\bf V}_j^H {\bf H}_{ij} \right] {\bf U}_i\right)}_{f_i({\bf U}_i,{\bf V}_{j\neq i})} + \sum_{j=1}^K \underbrace{ \tr\left( {\bf V}_j^H \left[ \sum_{i\neq j} {\bf H}_{ij}^H {\bf U}_i {\bf U}_i^H {\bf H}_{ij}^H \right] {\bf V}_j\right)}_{g_j({\bf V}_j,{\bf U}_{i\neq j})} =0.
\end{eqnarray}
In the above, $(a)$ follows from the fact that ${\bf X}={\bf 0} \Leftrightarrow  \tr\left({\bf X}{\bf X}^H\right)=0$ and the sum is zero iff all the (non-negative) summands are zero; and (b) holds because $\sum_{i=1}^K f_i({\bf U}_i,{\bf V}_{\sim i})=\sum_{j=1}^K  g_j({\bf V}_j,{\bf U}_{\sim j})$.
Using the notations introduced above, where $f_i$ and $g_j$ are non-negative real functions, we note that \eqref{IAcondition} admits the same solution set as the optimization problem
\begin{equation}
\min_{\scriptsize \begin{array}{c}({\bf U}_1,\ldots, {\bf U}_K)\in V_{N,d}^K,\\ ({\bf V}_1,\ldots, {\bf V}_K)\in V_{M,d}^K \end{array}}  \sum_{i=1}^K f_i({\bf U}_i,{\bf V}_{\sim i}) + \sum_{j=1}^K g_j({\bf V}_j,{\bf U}_{\sim j}). \label{opt_formulation}
\end{equation}
We propose to solve \eqref{opt_formulation} via message passing, specifically using the min-sum algorithm \cite{Yedidia_message_passing_inference_and_optimization_2011}. 
In order to do that, we construct a graph with $2K$ variable nodes ${\bf U}_i$, $i=1\ldots K$ and ${\bf V}_j$, $j=1\ldots K$, and $2K$ function nodes $f_i$, $i=1\ldots K$ and $g_j$, $j=1\ldots K$. A connection in the graph represents dependency of a function on the corresponding variable. An example of such a graph obtained for a 3-user network is shown in Fig.~\ref{fig_3user_graph}.\\

Note that according to the above graph construction, $f_i$ depends on ${\bf U}_i$ and on all ${\bf V}_j, \, j\neq i$; this assumption is valid for fully connected networks (${\bf H}_{ij}\neq {\bf 0} \  \forall (i,j)$). For partially connected networks (see \cite{Ruan_Lau_IA_PC_channels_TSP2011,Guillaud_Gesbert_Eusipco2011}), the connectivity of the graph should be adjusted accordingly, i.e. if ${\bf H}_{ij}={\bf 0}$, the edges between $f_i$ and ${\bf V}_j$, as well as between $g_j$ and ${\bf U}_i$ should be removed.\\

\section{Message-Passing Solution}
\label{section_MP}
In this section, we consider the application of the min-sum algorithm to the problem outlined above, i.e. we assume that messages are exchanged between the nodes in the graph. We now give the explicit formulation of the messages in our case of interest.

\subsection{Message computation}
\label{section_message_computation}

\subsubsection{Variable-to-function messages}
The general form of the variable-to-function updates in the considered message-passing algorithm can be computed as \cite{Yedidia_message_passing_inference_and_optimization_2011}
\begin{equation}
m_{a\rightarrow b}(x_a) = \sum_{i\in\mathcal{N}(a)\backslash b} m_{i\rightarrow a}(x_a),
\end{equation}
where $a$ is the variable node (and $x_a$ the corresponding variable), $b$ represents the function node, and $\mathcal{N}(a)$ denotes the set of neighbors of $a$.

For the considered problem, considering for instance the messages originating from ${\bf U}_i$ (and the set of its neighbors $\mathcal{N}({\bf U}_i)=\{f_i, g_{\sim i}\}$), %the message will take a different form depending on whether $i$ is a transmit precoder or a receive filter; namely, since $\mathcal{N}(f_i)=\{{\bf U}_i, {\bf V}_{\sim i}\}$ and $\mathcal{N}({\bf U}_i)=\{f_i, g_{\sim i}\}$, 
we must distinguish between messages going to $f_i$ and those going to one of the $g_j, j\neq i$:
\begin{equation}
m_{{\bf U}_i\rightarrow f_i}({\bf U}_i) = \sum_{j\neq i} m_{g_j\rightarrow {\bf U}_i}({\bf U}_i) \label{message_Ui_fi}
\end{equation}
while
\begin{equation}
m_{{\bf U}_i\rightarrow g_j}({\bf U}_i) = m_{f_i\rightarrow {\bf U}_i}({\bf U}_i) + \sum_{k\neq i,j} m_{g_k\rightarrow {\bf U}_i}({\bf U}_i) \label{message_Ui_gj}
\end{equation}
(note that, by a slight abuse of notation, we use the same name for the nodes in the graph and the matrices or functions they represent). By symmetry, we have
\begin{equation}
m_{{\bf V}_j\rightarrow g_j}({\bf V}_j) = \sum_{i\neq j} m_{f_i\rightarrow {\bf V}_j}({\bf V}_j)
\end{equation}
and
\begin{equation}
m_{{\bf V}_j\rightarrow f_i}({\bf V}_j) = m_{g_j\rightarrow {\bf V}_j}({\bf V}_j) + \sum_{k\neq i,j} m_{f_k\rightarrow {\bf V}_j}({\bf V}_j). \label{message_Vj_fi}
\end{equation}

\subsubsection{Function-to-variable messages}

Updates from function node $b$ to a variable node $a\in \mathcal{N}(b)$ take the general form
\begin{equation}
m_{b\rightarrow a}(x_a) = \min_{X_b \backslash x_a} \left[ C_b(X_b) + \sum_{j\in \mathcal{N}(b)\backslash a} m_{j\rightarrow b}(x_j) \right] \label{var2node}
\end{equation}
where $X_b$ is the set of all variables which are arguments of $C_b$.
Here again, considering a function node $f_i$ and its set of neighbors $\mathcal{N}(f_i)=\{{\bf U}_i, {\bf V}_{\sim i}\}$, we particularize  \eqref{var2node} depending on whether the message is going to ${\bf U}_i$ or to one of the ${\bf V}_j, j\neq i$:
\begin{eqnarray}
 \!\!\!\!\!\!\!\! m_{f_i\rightarrow {\bf U}_i}({\bf U}_i) \!\!\!\! &=& \!\!\!\! \min_{{\bf V}_{\sim i}} \Big[ f_i({\bf U}_i,{\bf V}_{\sim i}) + \sum_{j\neq i} m_{{\bf V}_j\rightarrow f_i}({\bf V}_j) \Big] \label{msg_fi_Ui} \\
\!\!\!\! \!\!\!\! m_{f_i\rightarrow {\bf V}_j}({\bf V}_j) \!\!\!\! &=& \!\!\!\!\min_{{\bf U}_i,{\bf V}_{\sim i,j}} \Big[ f_i({\bf U}_i,{\bf V}_{\sim i}) + m_{{\bf U}_i\rightarrow f_i}({\bf U}_i) \nonumber  \\
&&  + \sum_{k\neq i,j} m_{{\bf V}_k\rightarrow f_i}({\bf V}_k)\Big]. \label{msg_fi_Vj}
\end{eqnarray}
Again, by symmetry, we have
\begin{eqnarray}
 \!\!\!\!\!\!\!\! m_{g_j\rightarrow {\bf V}_j}({\bf V}_j) \!\!\!\!&= & \!\!\!\! \min_{{\bf U}_{\sim j}} \Big[ g_j({\bf V}_j,{\bf U}_{\sim j}) + \sum_{i\neq j} m_{{\bf U}_i\rightarrow g_j}({\bf U}_i)\Big] \label{msg_gj_Vj} \\
 \!\!\!\!\!\!\!\! m_{g_j\rightarrow {\bf U}_i}({\bf U}_i) \!\!\!\! &=&  \!\!\!\! \min_{{\bf V}_j,{\bf U}_{\sim i,j}} \Big[ g_j({\bf V}_j,{\bf U}_{\sim j}) + m_{{\bf V}_j\rightarrow g_j}({\bf V}_j) \nonumber \\
 &&   + \sum_{k\neq i,j} m_{{\bf U}_k\rightarrow g_j}({\bf U}_k)\Big]. \label{msg_gj_Ui}
\end{eqnarray}

\subsection{Convergence}
When considering tree-like graphs, propagating messages from the leaves to the root ensures that an exact solution is found with a finite number of message exchanges. Clearly, this is not the case here, since the considered graphs have loops -- in that case, the messages are initialized with random values, and message-passing is executed until convergence. In that case, the schedule (the order in which messages are propagated in the graph) can be arbitrary.
Although convergence is not always provable (see \cite{Moallemi_MinSum_convergence_IT10} for some convergence results applying to the min-sum algorithm), this will not constitute a problem here, as the proposed algorithm has been observed to converge reliably, as will be seen in Section~\ref{section_simul}.

After convergence, the variables of interest are computed from the local beliefs at each variable node, according to 
\begin{eqnarray}
{\bf U}_i^* &=& \arg\min_{{\bf U}_i} \sum_{a\in \mathcal{N}({\bf U}_i)} m_{a\rightarrow {\bf U}_i}({\bf U}_i), \\
{\bf V}_j^* &=& \arg\min_{{\bf V}_j} \sum_{a\in \mathcal{N}({\bf V}_j)} m_{a\rightarrow {\bf V}_j}({\bf V}_j).
\end{eqnarray}

Note that the messages considered here are in fact functions: in \eqref{msg_fi_Ui} for example, the message consists in the value of $m_{f_i\rightarrow {\bf U}_i}$ evaluated at all possible ${\bf U}_i$. If ${\bf U}_i$ takes on a finite number of values, it can be practical to compute the message by solving the minimization problem in the right-hand side for each possible value of ${\bf U}_i$. However, this method is clearly impractical for continuous variables. In the sequel, we turn our attention to the continuous case.

\section{Continuous variables case}
\label{section_continuouscase}
In order to have a compact (and computationally manageable) representation of continuous functions, we introduce a parameterization of our messages. Let us assume that every message  $m_{a\rightarrow b}({\bf X})$, where ${\bf X}$ is a truncated unitary matrix, takes the form
\begin{equation}
m_{a\rightarrow b}({\bf X}) = \tr\left({\bf X}^H {\bf Q}_{a\rightarrow b} {\bf X} \right) \label{eq_msgparameterization}
\end{equation}
for some positive semidefinite matrix ${\bf Q}_{a\rightarrow b}$. Clearly, under this assumption, the message $m_{a\rightarrow b}$ is equivalently represented by the corresponding matrix ${\bf Q}_{a\rightarrow b}$.
Using this parameterization, and resorting to approximations where necessary, we now show that the message computations from Section~\ref{section_message_computation} admit closed-form expressions of the form \eqref{eq_msgparameterization}.\\

Starting with \eqref{message_Ui_fi}, we notice that it can be transformed as
$m_{{\bf U}_i\rightarrow f_i}({\bf U}_i) = \sum_{j\neq i} \tr\left({\bf U}_i^H {\bf Q}_{g_j\rightarrow {\bf U}_i} {\bf U}_i \right) =  \tr\left({\bf U}_i^H \left[ \sum_{j\neq i} {\bf Q}_{g_j\rightarrow {\bf U}_i}\right] {\bf U}_i \right)$.
Identifying this expression with \eqref{eq_msgparameterization} shows that ${\bf Q}_{{\bf U}_i\rightarrow f_i}=\sum_{j\neq i} {\bf Q}_{g_j\rightarrow {\bf U}_i}$. Using a similar reasoning, \eqref{message_Ui_gj}--\eqref{message_Vj_fi} yield trivially ${\bf Q}_{{\bf U}_i\rightarrow g_j}={\bf Q}_{f_i\rightarrow {\bf U}_i}+\sum_{k\neq i,j} {\bf Q}_{g_k\rightarrow {\bf U}_i}$,  ${\bf Q}_{{\bf V}_j\rightarrow g_j}=\sum_{i\neq j} {\bf Q}_{f_i\rightarrow {\bf V}_j}$ and ${\bf Q}_{{\bf V}_j\rightarrow f_i}= {\bf Q}_{g_j\rightarrow {\bf V}_j} + \sum_{k\neq i,j}{\bf Q}_{f_k\rightarrow {\bf V}_j}$.\\

The case of the the function-to-variable messages is more interesting, since we now seek the closed-form expressions of the solutions to the minimization problems \eqref{msg_fi_Ui}--\eqref{msg_gj_Ui}. Let us first turn our attention to \eqref{msg_fi_Ui}, and notice that
{\small \begin{eqnarray}
&&\!\!\!\!\!\!\!\!\!\!\!\!\!\!\!\!\!\! f_i({\bf U}_i,{\bf V}_{\sim i}) + \sum_{j\neq i} m_{{\bf V}_j\rightarrow f_i}({\bf V}_j) \nonumber \\
%\!\!\!\!\!\!\!\!\!\!&=& \!\!\!\!\! \tr\bigg( {\bf U}_i^H \Big[ \sum_{j\neq i} {\bf H}_{ij} {\bf V}_j {\bf V}_j^H {\bf H}_{ij}^H \Big] {\bf U}_i\bigg) + \sum_{j\neq i}  \tr\left({\bf V}_j^H {\bf Q}_{{\bf V}_j\rightarrow f_i} {\bf V}_j \right) \nonumber \\
&&\!\!\!\!\!\!\!\!\!\!\!\!\!\!\!\!\!\!= \sum_{j\neq i}\left[ \tr\left( {\bf U}_i^H {\bf H}_{ij} {\bf V}_j {\bf V}_j^H {\bf H}_{ij}^H  {\bf U}_i\right)+ \tr\left({\bf V}_j^H {\bf Q}_{{\bf V}_j\rightarrow f_i} {\bf V}_j \right) \right].
\end{eqnarray}}
Note that each term in the above sum depends only on a single ${\bf V}_j$. This indicates that the minimization over ${\bf V}_{\sim i}$ in \eqref{msg_fi_Ui} is separable, and therefore
{\small \begin{eqnarray}
&&\!\!\!\!\!\!\!\!\!\!\!\!\!\!\!\!\!\! m_{f_i\rightarrow {\bf U}_i}({\bf U}_i) \nonumber \\
&&\!\!\!\!\!\!\!\!\!\!\!\!\!\!\!\!\!\!= \sum_{j\neq i}   \min_{{\bf V}_j} \left[ \tr\left( {\bf U}_i^H {\bf H}_{ij} {\bf V}_j {\bf V}_j^H {\bf H}_{ij}^H  {\bf U}_i\right)+ \tr\left({\bf V}_j^H {\bf Q}_{{\bf V}_j\rightarrow f_i} {\bf V}_j \right)\right] \nonumber \\
&&\!\!\!\!\!\!\!\!\!\!\!\!\!\!\!\!\!\!= \sum_{j\neq i}   \min_{{\bf V}_j} \tr \left( {\bf V}_j^H \left[ {\bf H}_{ij}^H  {\bf U}_i {\bf U}_i^H {\bf H}_{ij} + {\bf Q}_{{\bf V}_j\rightarrow f_i} \right]{\bf V}_j \right).% \label{msg_fi_Ui}
\end{eqnarray}}
Here, we resort to our first approximation, and assume that $\arg\min \tr \left( {\bf V}_j^H \left[ {\bf H}_{ij}^H  {\bf U}_i {\bf U}_i^H {\bf H}_{ij} + {\bf Q}_{{\bf V}_j\rightarrow f_i} \right]{\bf V}_j \right) \approx \arg\min \tr \left( {\bf V}_j^H {\bf Q}_{{\bf V}_j\rightarrow f_i} {\bf V}_j \right)$. Note that this approximation becomes exact if ${\bf V}_j^H {\bf H}_{ij}^H {\bf U}_i={\bf 0}$, i.e. in particular at convergence, when \eqref{IAcondition} is fulfilled.
Letting ${\bf V}_j^0 = \numin \left( {\bf Q}_{{\bf V}_j\rightarrow f_i} \right) $ for all $j\neq i$, we have 
{\small 
\begin{eqnarray}
&& \!\!\!\!\!\!\!\!\!\!\!\!\!\!\! m_{f_i\rightarrow {\bf U}_i}({\bf U}_i) \nonumber \\
%&&= \sum_{j\neq i}   \tr \left( {{\bf V}_j^0}^H \left[ {\bf H}_{ij}^H  {\bf U}_i {\bf U}_i^H {\bf H}_{ij} + {\bf Q}_{{\bf V}_j\rightarrow f_i} \right]{\bf V}_j^0 \right) \\% \label{msg_fi_Ui}
&&\!\!\!\!\!\!\!\!\!\!\!\!\!\!\!= \sum_{j\neq i}   \tr \left( {\bf U}_i^H {\bf H}_{ij} {\bf V}_j^0 {{\bf V}_j^0}^H  {\bf H}_{ij}^H  {\bf U}_i \right)+ \tr \left({{\bf V}_j^0}^H{\bf Q}_{{\bf V}_j\rightarrow f_i} {\bf V}_j^0 \right) \\
&&\!\!\!\!\!\!\!\!\!\!\!\!\!\!\!=   \tr \bigg( {\bf U}_i^H \Big[ \sum_{j\neq i} {\bf H}_{ij} {\bf V}_j^0 {{\bf V}_j^0}^H  {\bf H}_{ij}^H +\I \frac{1}{d}\tr \left({{\bf V}_j^0}^H{\bf Q}_{{\bf V}_j\rightarrow f_i} {\bf V}_j^0 \right) \Big] {\bf U}_i  \bigg). \nonumber
\end{eqnarray}}
This yields the final message computation rule  ${\bf Q}_{f_i\rightarrow {\bf U}_i}= \sum_{j\neq i} {\bf H}_{ij} {\bf V}_j^0 {{\bf V}_j^0}^H  {\bf H}_{ij}^H +\I \frac{1}{d}\tr \left({{\bf V}_j^0}^H{\bf Q}_{{\bf V}_j\rightarrow f_i} {\bf V}_j^0 \right)$.\\

The case of \eqref{msg_fi_Vj} follows a similar derivation:
{\small
\begin{eqnarray}
&& \!\!\!\!\!\!\!\!\!\!\!\!\!\!\! m_{f_i\rightarrow {\bf V}_j}({\bf V}_j) \nonumber \\
&&\!\!\!\!\!\!\!\!\!\!\!\!\!\!\!= \min_{{\bf U}_i,{\bf V}_{\sim i,j}} \Bigg[ \tr\bigg( {\bf U}_i^H \Big[ \sum_{k\neq i} {\bf H}_{ik} {\bf V}_k {\bf V}_k^H {\bf H}_{ik}^H \Big] {\bf U}_i\bigg) \\
&& + \tr\left({\bf U}_i^H {\bf Q}_{{\bf U}_i\rightarrow f_i} {\bf U}_i \right)  + \sum_{k\neq i,j} \tr\left({\bf V}_k^H {\bf Q}_{{\bf V}_k\rightarrow f_i} {\bf V}_k \right) \Bigg] \nonumber \\
&&\!\!\!\!\!\!\!\!\!\!\!\!\!\!\!= \min_{{\bf U}_i,{\bf V}_{\sim i,j}} \Bigg[ \tr\left( {\bf U}_i^H {\bf H}_{ij} {\bf V}_j {\bf V}_j^H {\bf H}_{ij}^H {\bf U}_i\right) \nonumber\\
&& +  \tr\bigg( {\bf U}_i^H \underbrace{\Big[{\bf Q}_{{\bf U}_i\rightarrow f_i}+ \frac{1}{2}\sum_{k\neq i,j} {\bf H}_{ik} {\bf V}_k {\bf V}_k^H {\bf H}_{ik}^H \Big]}_{{\bf R}} {\bf U}_i\bigg)  \\ 
&&+ \sum_{k\neq i,j} \tr\bigg({\bf V}_k^H \underbrace{\Big[ {\bf Q}_{{\bf V}_k\rightarrow f_i} + \frac{1}{2} {\bf H}_{ik}^H {\bf U}_i {\bf U}_i^H {\bf H}_{ik} \Big]}_{{\bf S}_k}{\bf V}_k \bigg) \Bigg].\nonumber
\end{eqnarray}}
Again, we choose to approximate the minimization above by assuming that $\arg \min_{{\bf U}_i,{\bf V}_{\sim i,j}}  \Big[\tr\left( {\bf U}_i^H {\bf H}_{ij} {\bf V}_j {\bf V}_j^H {\bf H}_{ij}^H {\bf U}_i\right) +  \tr\left( {\bf U}_i^H {\bf R}{\bf U}_i\right) +  \sum_{k\neq i,j} \tr\left({\bf V}_k^H {\bf S}_k {\bf V}_k \right) \Big]\approx \arg \min_{{\bf U}_i,{\bf V}_{\sim i,j}} \Big[ \tr\left( {\bf U}_i^H {\bf R}{\bf U}_i\right) +  \sum_{k\neq i,j} \tr\left({\bf V}_k^H {\bf S}_k {\bf V}_k \right) \Big]$. Note however that $\bf R$ depends on ${\bf V}_{\sim i,j}$ and ${\bf S}_k$ is a function of ${\bf U}_i$, so the resulting minimization problem is not separable. We resort to alternating minimization of $\tr\left( {\bf U}_i^H {\bf R}{\bf U}_i\right)$ and the $\tr\left({\bf V}_k^H {\bf S}_k {\bf V}_k \right)$; starting from an arbitrary ${\bf U}_i^{(0)}\in V_{N,d}$, we apply the following update rules at iteration $n$:
{\small 
\begin{eqnarray}
\!\!\!\!\!\!\!\!\!\!\!\! {\bf V}_k^{(n+1)} \!\!\!\!\!\!&=&\!\!\!\!\!\! \numin \Big(  {\bf Q}_{{\bf V}_k\rightarrow f_i} + \frac{1}{2} {\bf H}_{ik}^H {\bf U}_i^{(n)} {{\bf U}_i^{(n)}}^H {\bf H}_{ik}  \Big),    \forall k\neq i,j  \label{iter1} \\
\!\!\!\!\!\!\!\!\!\!\!\! {\bf U}_i^{(n+1)} \!\!\!\!\!\!&=& \!\!\!\!\!\!\numin \Big( {\bf Q}_{{\bf U}_i\rightarrow f_i}+ \frac{1}{2}\sum_{k\neq i,j} {\bf H}_{ik} {\bf V}_k^{(n+1)} {{\bf V}_k^{(n+1)}}^H {\bf H}_{ik}^H \Big). \label{iter2}
\end{eqnarray}}
Clearly, when $n\rightarrow \infty$, the objective function $\tr\left( {\bf U}_i^H {\bf R}{\bf U}_i\right) +  \sum_{k\neq i,j} \tr\left({\bf V}_k^H {\bf S}_k {\bf V}_k \right)$ converges (it is non-negative and non-increasing at each iteration). We can not provide any proof of optimality for this approach; nonetheless, experimental results obtained via this technique have been satisfactory. Letting ${\bf U}_i^{*},{\bf V}_{k}^{*}$ denote the convergence points of the iterations of \eqref{iter1}--\eqref{iter2}, we finally obtain
{\small 
\begin{eqnarray}
&&\!\!\!\!\!\!\!\!\!\!\!\! m_{f_i\rightarrow {\bf V}_j}({\bf V}_j) \nonumber \\
&&\!\!\!\!\!\!\!\!\!\!\!\! =  \tr\left( {\bf V}_j^H {\bf H}_{ij}^H {\bf U}_i^* {{\bf U}_i^*}^H {\bf H}_{ij} {\bf V}_j\right) \!+\!\!\!\sum_{k\neq i,j}\tr\left( {{\bf U}_i^*}^H {\bf H}_{ik} {\bf V}_k^* {{\bf V}_k^*}^H {\bf H}_{ik}^H {\bf U}_i^* \right) \nonumber  \\
&&+ \tr\left({{\bf U}_i^*}^H {\bf Q}_{{\bf U}_i\rightarrow f_i} {\bf U}_i^* \right)  + \sum_{k\neq i,j} \tr\left({{\bf V}_k^*}^H {\bf Q}_{{\bf V}_k\rightarrow f_i} {\bf V}_k^* \right),
\end{eqnarray}}
which corresponds to the following rule: 
{\small
\begin{eqnarray}
&&\!\!\!\!\!\!\!\!\!\!\!\!{\bf Q}_{f_i\rightarrow {\bf V}_j} \nonumber  \\
&&\!\!\!\!\!\!\!\!\!\!\!\!=  {\bf H}_{ij}^H {\bf U}_i^* {{\bf U}_i^*}^H {\bf H}_{ij} + \I \frac{1}{d} \bigg[\sum_{k\neq i,j}\tr\Big( {{\bf U}_i^*}^H {\bf H}_{ik} {\bf V}_k^* {{\bf V}_k^*}^H {\bf H}_{ik}^H {\bf U}_i^* \Big) \nonumber \\
&&\!\!\!+ \tr\left({{\bf U}_i^*}^H {\bf Q}_{{\bf U}_i\rightarrow f_i} {\bf U}_i^* \right)  + \sum_{k\neq i,j} \tr\Big({{\bf V}_k^*}^H {\bf Q}_{{\bf V}_k\rightarrow f_i} {\bf V}_k^* \Big) \bigg].
\end{eqnarray}}
Note that the terms proportional to the identity matrix that appear in ${\bf Q}_{f_i\rightarrow {\bf U}_i}$ or ${\bf Q}_{f_i\rightarrow {\bf V}_j}$ above only add a constant (independent of the considered variable) to the objective function. As such, these terms can be omitted from the message-passing implementation.
The matrices corresponding to the messages in \eqref{msg_gj_Vj} and \eqref{msg_gj_Ui} are obtained in a similar manner.

We summarize below the message computation rules using the parametric form, which form the proposed message-passing IA (MPIA) algorithm:
{\small 
\begin{eqnarray}
{\bf Q}_{{\bf U}_i\rightarrow f_i}&\!\!\!\!\!=&\!\!\!\!\!\sum_{j\neq i} {\bf Q}_{g_j\rightarrow {\bf U}_i}\\
{\bf Q}_{{\bf U}_i\rightarrow g_j}&\!\!\!\!\!=&\!\!\!\!\!{\bf Q}_{f_i\rightarrow {\bf U}_i}+\sum_{k\neq i,j} {\bf Q}_{g_k\rightarrow {\bf U}_i} \\
{\bf Q}_{{\bf V}_j\rightarrow g_j}&\!\!\!\!\!=&\!\!\!\!\!\sum_{i\neq j} {\bf Q}_{f_i\rightarrow {\bf V}_j}\\
{\bf Q}_{{\bf V}_j\rightarrow f_i}&\!\!\!\!\!=& \!\!\!\!\!{\bf Q}_{g_j\rightarrow {\bf V}_j} + \sum_{k\neq i,j}{\bf Q}_{f_k\rightarrow {\bf V}_j} \label{Q_Vj_fi}  \\
{\bf Q}_{f_i\rightarrow {\bf U}_i}&\!\!\!\!\!=&\!\!\!\!\! \sum_{j\neq i} {\bf H}_{ij} {\bf V}_j^0 {{\bf V}_j^0}^H  {\bf H}_{ij}^H, \nonumber \\
&& \mathrm{where} \quad {\bf V}_j^0 = \numin \left( {\bf Q}_{{\bf V}_j\rightarrow f_i} \right) \ \forall j\neq i    \label{Q_fi_Ui}   \\
{\bf Q}_{f_i\rightarrow {\bf V}_j} &\!\!\!\!\!=&\!\!\!\!\!  {\bf H}_{ij}^H {\bf U}_i^* {{\bf U}_i^*}^H {\bf H}_{ij}   \label{Q_fi_Vj} \\
{\bf Q}_{g_j\rightarrow {\bf V}_j}&\!\!\!\!\!=& \!\!\!\!\!\sum_{i\neq j} {\bf H}_{ij}^H {\bf U}_i^0 {{\bf U}_i^0}^H  {\bf H}_{ij}, \nonumber \\
&& \mathrm{where} \quad {\bf U}_i^0 = \numin \left( {\bf Q}_{{\bf U}_i\rightarrow g_j} \right) \ \forall i\neq j  \label{Q_gj_Vj}  \\
{\bf Q}_{g_j\rightarrow {\bf U}_i} &\!\!\!\!\!=& \!\!\!\!\! {\bf H}_{ij} {\bf V}_j^* {{\bf V}_j^*}^H {\bf H}_{ij}^H  \label{Q_gj_Ui}.
\end{eqnarray}}
In \eqref{Q_fi_Vj}, ${\bf U}_i^*$ is computed by iterating \eqref{iter1}--\eqref{iter2}, while in \eqref{Q_gj_Ui}, ${\bf V}_j^*$ is obtained by iterating
{\small
\begin{eqnarray}
\!\!\!\!\!\!\!\!\!\!\!\! {\bf U}_k^{(n+1)}  \!\!\!\!\!\!&=& \!\!\!\!\!\! \numin \Big(  {\bf Q}_{{\bf U}_k\rightarrow g_j} + \frac{1}{2} {\bf H}_{kj} {\bf V}_j^{(n)} {{\bf V}_j^{(n)}}^H {\bf H}_{kj}^H  \Big),  \ \forall k\neq i,j  \\
\!\!\!\!\!\!\!\!\!\!\!\! {\bf V}_j^{(n+1)} \!\!\!\!\!\! &=& \!\!\!\!\!\! \numin \Big( {\bf Q}_{{\bf V}_j\rightarrow g_j}+ \frac{1}{2}\sum_{k\neq i,j} {\bf H}_{kj}^H {\bf U}_k^{(n+1)} {{\bf U}_k^{(n+1)}}^H {\bf H}_{kj} \Big)
\end{eqnarray}}
after initialization with an arbitrary ${\bf V}_j^{(0)}$.\\

\section{Link with the iterative leakage minimization algorithm}
\label{section_ILM}

We now point out some ties between the MPIA algorithm for the continuous case described in Section~\ref{section_continuouscase} and the iterative leakage minimization (ILM) algorithm \cite[Algorithm~1]{Gomadamit08}; for simplicity, in this section, we will consider only centralized implementations of both MPIA and ILM. It can be shown that ILM is a particular case of MPIA, obtained for a certain schedule. To see this, assume that all messages in the MPIA are initialized with zero matrices, and propagate the messages according to the following schedule: $m_{g_j\rightarrow {\bf V}_j} \ \forall j$, $m_{{\bf V}_j\rightarrow f_i}  \ \forall i\neq j$, $m_{f_i\rightarrow {\bf U}_i} \ \forall i$, and $m_{{\bf U}_i\rightarrow g_j} \ \forall j\neq i$. The other messages are never updated, and therefore ${\bf Q}_{{\bf U}_i\rightarrow f_i}$, ${\bf Q}_{{\bf V}_j\rightarrow g_j}$, ${\bf Q}_{f_i\rightarrow {\bf V}_j}$ and ${\bf Q}_{g_j\rightarrow {\bf U}_i}$ remain at their initial value $\bf 0$ throughout the algorithm.

In order to see the correspondence, consider the first computation of \eqref{Q_gj_Vj}. Since ${\bf Q}_{{\bf U}_i\rightarrow g_j}={\bf 0}$, the $\numin$ operator returns random, isotropically distributed matrices ${\bf U}_i^0$, and therefore ${\bf Q}_{g_j\rightarrow {\bf V}_j}$ is random. \eqref{Q_Vj_fi} yields ${\bf Q}_{{\bf V}_j\rightarrow f_i}= {\bf Q}_{g_j\rightarrow {\bf V}_j} \forall i$ since all other terms in the sum are zero. Next, considering \eqref{Q_fi_Ui}, the ${\bf V}_j^0$ computed as $\numin \left( {\bf Q}_{{\bf V}_j\rightarrow f_i} \right)$ correspond to the random initialization of the transmit precoders ${\bf V}^{[j]}$ in \cite[Algorithm~1]{Gomadamit08}.
From this point on, it is easy to prove by induction that the two algorithms perform the same computations since
(using the notations ${\bf V}^{[j]}$ and ${\bf U}^{[i]}$ for the precoders and receive filters from \cite{Gomadamit08}):
{\small 
\begin{eqnarray}
(m_{f_i\rightarrow {\bf U}_i}) &&  \!\!\!\!\!\!\!\! \left\{ \begin{array}{l}
{\bf V}_j^0 = \numin \left( \sum_{i\neq j} {\bf H}_{ij}^H {\bf U}^{[i]} {{\bf U}^{[i]}}^H  {\bf H}_{ij} \right) = {\bf V}^{[j]}\\
{\bf Q}_{f_i\rightarrow {\bf U}_i}= \sum_{j\neq i} {\bf H}_{ij} {\bf V}^{[j]} {{\bf V}^{[j]}}^H  {\bf H}_{ij}^H 
\end{array} \right. \\
(m_{{\bf U}_i\rightarrow g_j}) && \!\!\!\!\!\!\!\!{\bf Q}_{{\bf U}_i\rightarrow g_j}={\bf Q}_{f_i\rightarrow {\bf U}_i} \\
(m_{g_j\rightarrow {\bf V}_j}) &&\!\!\!\!\!\!\!\!\left\{  \begin{array}{l}
{\bf U}_i^0 = \numin \left( \sum_{j\neq i} {\bf H}_{ij} {\bf V}^{[j]} {{\bf V}^{[j]}}^H  {\bf H}_{ij}^H \right) = {\bf U}^{[i]} \\
{\bf Q}_{g_j\rightarrow {\bf V}_j}= \sum_{i\neq j} {\bf H}_{ij}^H {\bf U}^{[i]} {{\bf U}^{[i]}}^H  {\bf H}_{ij}
\end{array} \right.\\
(m_{{\bf V}_j\rightarrow f_i}) && \!\!\!\!\!\!\!\!{\bf Q}_{{\bf V}_j\rightarrow f_i}= {\bf Q}_{g_j\rightarrow {\bf V}_j}.
\end{eqnarray} }

\section{Simulation Results}
\label{section_simul}

In this section, the proposed algorithm is validated using numerical simulations, and compared to the ILM algorithm from \cite{Gomadamit08}. All simulations presented here are for a 3-user IC, with $4\times 4$ MIMO channels and $d=2$ streams per user. A regular schedule was used for the MPIA algorithm, whereby the messages are propagated in the following order: $m_{g_j\rightarrow {\bf V}_j}$, $m_{f_i\rightarrow {\bf V}_j}$, $m_{{\bf V}_j\rightarrow f_i}$, $m_{{\bf V}_j\rightarrow g_j}$, $m_{f_i\rightarrow {\bf U}_i}$, $m_{g_j\rightarrow {\bf U}_i}$, $m_{{\bf U}_i\rightarrow g_j}$, $m_{{\bf U}_i\rightarrow f_i}$ (this sequence constituting one iteration). For the ILM algorithm, one iteration consists in $m_{g_j\rightarrow {\bf V}_j}$, $m_{{\bf V}_j\rightarrow f_i}$, $m_{f_i\rightarrow {\bf U}_i}$, and $m_{{\bf U}_i\rightarrow g_j}$, as outlined in Section~\ref{section_ILM}.\\

\begin{figure}
  \centering
  \includegraphics[width=.95\columnwidth]{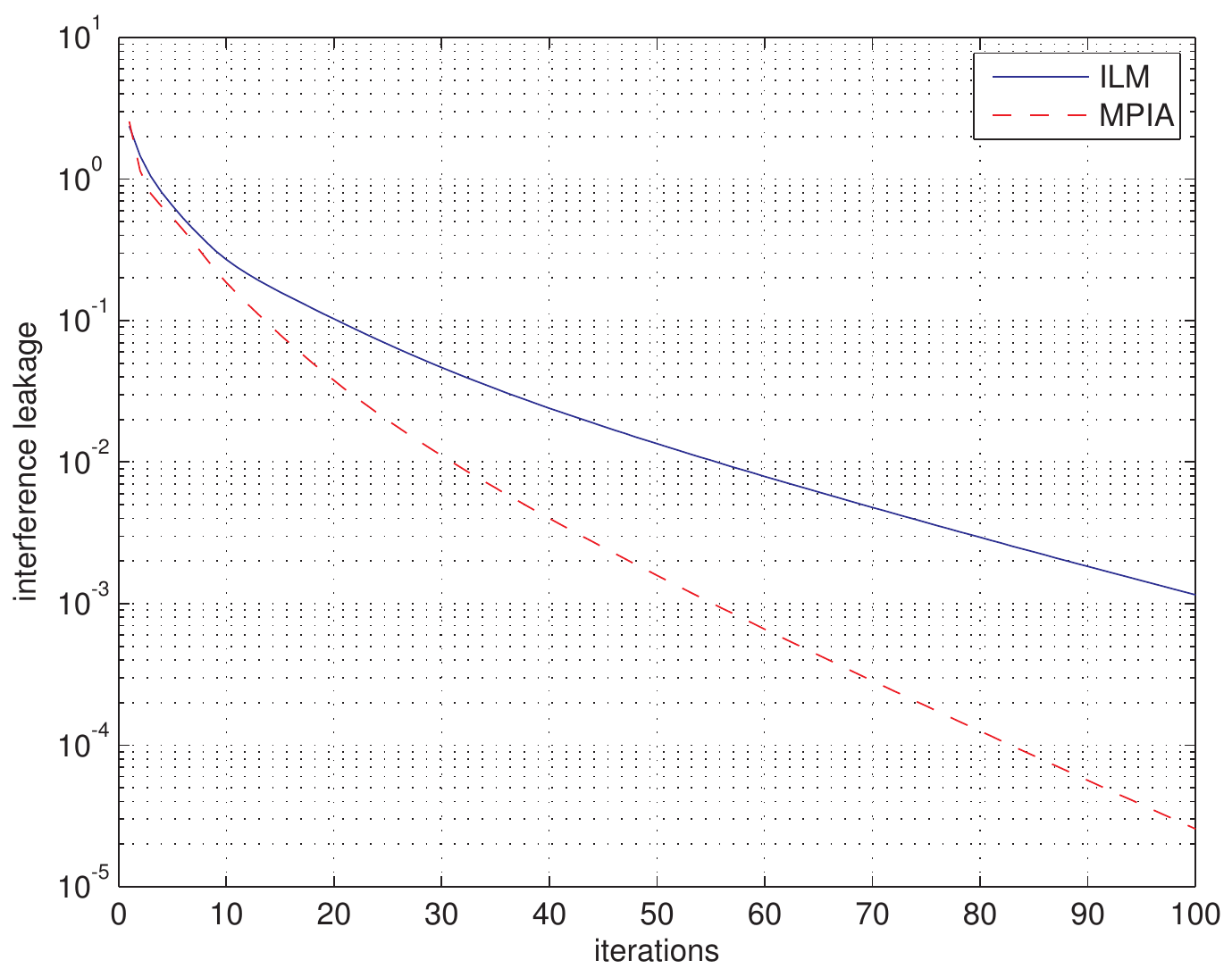}
  \caption{Convergence trajectory (leakage vs. iterations) of ILM and MPIA for one random channel realization}
  \label{fig_trajectory}
\end{figure}

We first compare the MPIA algorithm to the ILM in terms of convergence speed. 
Fig.~\ref{fig_trajectory} depicts the interference leakage -- the objective function from \eqref{opt_formulation} -- versus the number of iterations for both algorithms for a random channel realization (the same channel values were used for both algorithms). Observe that MPIA converges faster than ILM to zero leakage. Note that although the curves in Fig.~\ref{fig_trajectory} are related to one particular channel realization, this behavior is observed consistently for other channel values. 

\begin{figure}
  \centering
  \includegraphics[width=.95\columnwidth]{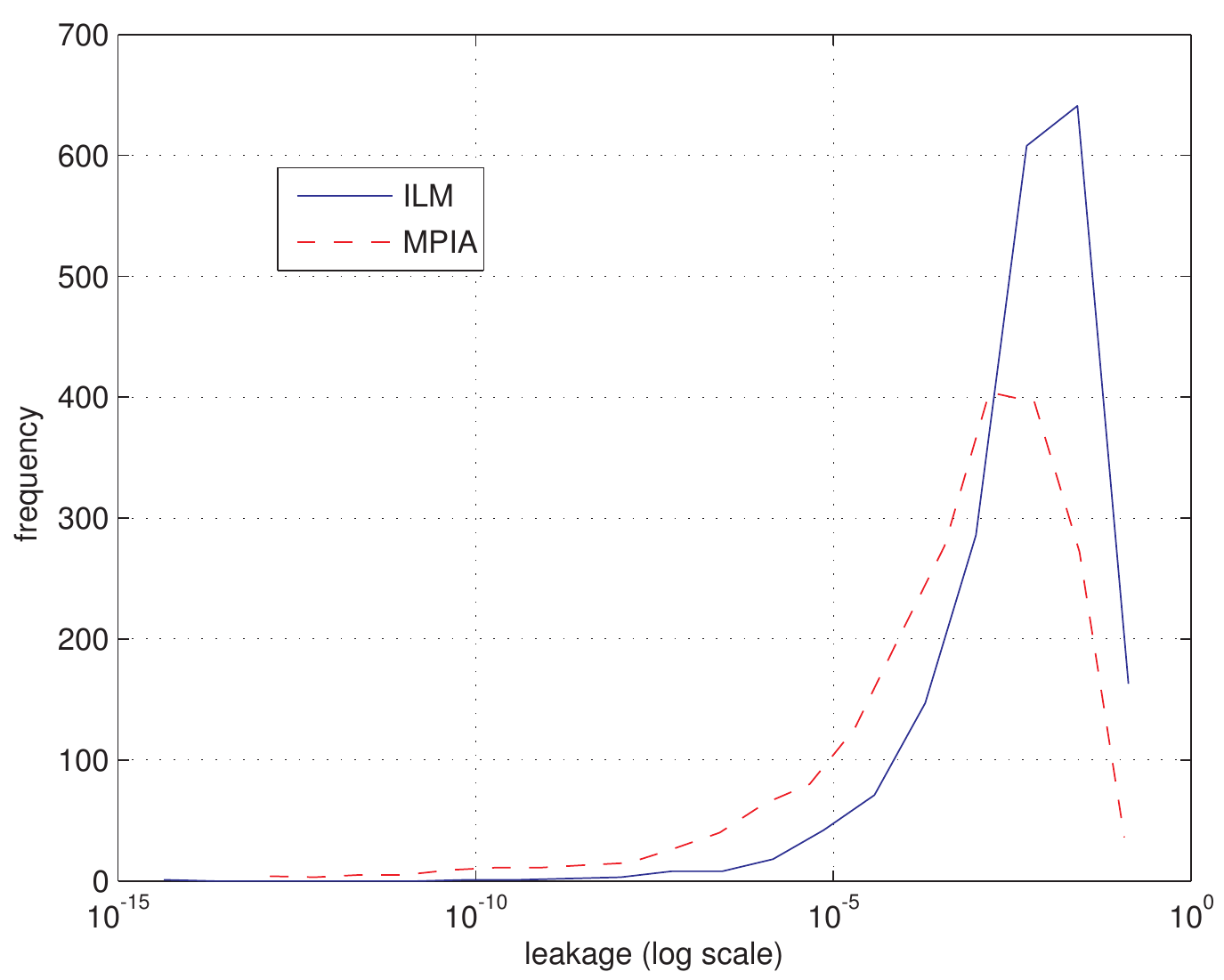}
  \caption{Empirical leakage distributions over 2000 Gaussian i.i.d. channels realizations, after 100 iterations of MPIA and ILM.}
  \label{fig_lk_distribution}
\end{figure}

In order to evaluate the respective accuracy of MPIA and ILM over a distribution of channels, we compare in Fig.~\ref{fig_lk_distribution} the statistics of the leakage achieved after 100 iterations of each algorithm, for channels with coefficients drawn i.i.d. from a complex circularly symmetric Gaussian distribution. The curves depict the empirical distribution of the leakage obtained by running both algorithms over 2000 channel realizations. These results show that, for a fixed number of iterations, MPIA achieves a lower leakage than ILM, the bulk of the leakage distribution being further left (towards lower values) for MPIA compared to ILM. The mean (computed in log-scale) leakage is  $4\cdot10^{-3}$ for ILM, and $3.8\cdot10^{-4}$ for MPIA.\\

Arguably, the faster convergence of MPIA should be mitigated by the fact that the complexity per iteration in ILM and in MPIA are not comparable. However, we argue in the next section that computational complexity is not the most relevant metric in the case of a distributed implementation.\\

\section{Distributed implementation}
\label{section_distributed}
A distributed implementation of the proposed algorithm can be obtained by distributing the computations associated with the nodes in the graph (Fig.~\ref{fig_3user_graph}) among the physical devices in the network. Clearly, this mapping will influence the amount of data exchanged between the devices (e.g. in wireless systems, exchanging data between devices is costly in terms of energy and bandwidth; on the other hand, it is reasonable to assume that messages exchanged on the graph between nodes on the same device do not incur any communication costs. In such systems, we expect communication costs to outweigh computational complexity). \\

An obvious mapping of the graph to the devices which preserves the symmetry between transmitters and receivers is obtained by associating ${\bf U}_i$ and $f_i$ to receiver $i$, while $g_j$ and ${\bf V}_j$ are associated to transmitter $j$. In that case, only $m_{f_i\rightarrow {\bf V}_j}$, $m_{{\bf V}_j\rightarrow f_i}$, $m_{g_j\rightarrow {\bf U}_i}$ and $m_{{\bf U}_i\rightarrow g_j}$ need to be transmitted over the air, while the computation and exchange of $m_{g_j\rightarrow {\bf V}_j}$, $m_{{\bf V}_j\rightarrow g_j}$, $m_{f_i\rightarrow {\bf U}_i}$ and $m_{{\bf U}_i\rightarrow f_i}$ is confined to a single device.

There is no guarantee however that this mapping is optimal in any sense. In fact, a meaningful evaluation of a distributed implementation should involve an analysis of the amount of data required to faithfully represent the messages (e.g. by considering the quantization of matrix ${\bf Q}_{a\rightarrow b}$ in \eqref{eq_msgparameterization}) and of the costs of communication between devices in the network (in some cases this would include considering out-of-band communications over e.g. a backhaul network, and potential optimizations enabled by the broadcast nature of the wireless medium). This analysis however is clearly outside of the scope of the present article. We envision this as a future research direction.\\

\section{Conclusion}
We have introduced an iterative solution to the problem of interference alignment over MIMO channels based on MP applied to a suitable graph. A parameterization of the messages that enables the use of this algorithm over continuous variable spaces was introduced, and closed-form approximations of the messages were derived. We have show that the iterative leakage minimization algorithm of Cadambe \emph{et al.} is a particular case of our message-passing algorithm. The proposed algorithm was shown to outperform ILM in terms of convergence speed. Finally, we discussed the distributed implementation of the proposed technique.\\

\section*{Acknowledgement}
This work was supported by the FP7 projects HIATUS (grant 265578) of the European Commission (EC) and by the Austrian Science Fund (FWF) through grant NFN SISE (S106). The authors also acknowledge the support of the EC Newcom\# Network of Excellence in Wireless Communications.\\

\bibliographystyle{IEEEtran}
%\begin{thebibliography}{1}

  %0.5em minus 0.4em\relax Harlow, England: Addison-Wesley, 1999.
\balance
%\end{thebibliography}
\bibliography{biblio.bib}

\begin{thebibliography}{10}
\providecommand{\url}[1]{#1}
\csname url@samestyle\endcsname
\providecommand{\newblock}{\relax}
\providecommand{\bibinfo}[2]{#2}
\providecommand{\BIBentrySTDinterwordspacing}{\spaceskip=0pt\relax}
\providecommand{\BIBentryALTinterwordstretchfactor}{4}
\providecommand{\BIBentryALTinterwordspacing}{\spaceskip=\fontdimen2\font plus
\BIBentryALTinterwordstretchfactor\fontdimen3\font minus
  \fontdimen4\font\relax}
\providecommand{\BIBforeignlanguage}[2]{{%
\expandafter\ifx\csname l@#1\endcsname\relax
\typeout{** WARNING: IEEEtran.bst: No hyphenation pattern has been}%
\typeout{** loaded for the language `#1'. Using the pattern for}%
\typeout{** the default language instead.}%
\else
\language=\csname l@#1\endcsname
\fi
#2}}
\providecommand{\BIBdecl}{\relax}
\BIBdecl

\bibitem{Gou_Jafar_DoF_MIMO_Kuser_IC_IT2010}
T.~Gou and S.~A. Jafar, ``Degrees of freedom of the {$K$} user {$M\times N$
  MIMO} interference channel,'' \emph{IEEE Transactions on Information Theory},
  vol.~56, no.~12, pp. 6040--6057, Dec. 2010.

\bibitem{Tresch_Guillaud_Riegler_SSP09}
R.~Tresch, M.~Guillaud, and E.~Riegler, ``On the achievability of interference
  alignment in the {K}-user constant {MIMO} interference channel,'' in
  \emph{Proc. IEEE Workshop on Statistical Signal Processing (SSP)}, Cardiff,
  U.K., Sep. 2009.

\bibitem{Gomadamit08}
K.~Gomadam, V.~R. Cadambe, and S.~A. Jafar, ``A distributed numerical approach
  to interference alignment and applications to wireless interference
  networks,'' \emph{IEEE Transactions on Information Theory}, vol.~57, no.~6,
  pp. 3309--3322, Jun. 2011.

\bibitem{Kschischang_sum_product_algorithm}
F.~R. Kschischang, B.~J. Frey, and H.-A. Loeliger, ``Factor graphs and the
  sum-product algorithm,'' \emph{IEEE Transactions on Information Theory},
  vol.~47, no.~2, pp. 498--519, Feb. 2001.

\bibitem{Aji_McEliece_generalized_distributive_law_it2000}
S.~M. Aji and R.~J. McEliece, ``The generalized distributive law,'' \emph{IEEE
  Transactions on Information Theory}, vol.~46, no.~2, pp. 325--343, Mar. 2000.

\bibitem{Walsh_Regalia_connecting_BP_with_ML}
J.~M. Walsh and P.~A. Regalia, ``Connecting belief propagation with maximum
  likelihood detection,'' in \emph{Proc. Int. Symposium on Turbo Codes and
  Related Topics}, Munich, Germany, Apr. 2006.

\bibitem{Yedidia_message_passing_inference_and_optimization_2011}
J.~S. Yedidia, ``Message-passing algorithms for inference and optimization,''
  \emph{Journal of Statistical Physics}, vol. 145, no.~4, pp. 860--890, Nov.
  2011.

\bibitem{Sohn_Lee_Andrews_BP_distributed_DL_beamforming_TWC2011}
I.~Sohn, S.~H. Lee, and J.~G. Andrews, ``Belief propagation for distributed
  downlink beamforming in cooperative {MIMO} cellular networks,'' \emph{IEEE
  Transactions on Wireless Communications}, vol.~10, no.~12, pp. 4140--4149,
  Dec. 2011.

\bibitem{GHC_distributed_approach_to_precoder_optimization_JASP2013}
I.~M. Guerreiro, D.~Hui, and C.~C. Cavalcante, ``A distributed approach to
  precoder selection using factor graphs for wireless communication networks,''
  \emph{EURASIP Journal on Applied Signal Processing}, vol.~83, 2013.

\bibitem{Schmidt_Utschick_Honig_Globecom2010}
D.~Schmidt, W.~Utschick, and M.~Honig, ``Large system performance of
  interference alignment in single-beam {MIMO} networks,'' in \emph{Proc. IEEE
  Global Telecommunications (Globecom) Conference}, Miami, FL, USA, Dec. 2010.

\bibitem{Yetis_Jafar_feasibility_conditions_IA_Globecom09}
C.~M. Yetis, S.~A. Jafar, and A.~H. Kayran, ``Feasibility conditions for
  interference alignment,'' in \emph{Proc. IEEE Globecom}, Honolulu, HI, USA,
  Nov. 2009.

\bibitem{Ruan_Lau_IA_PC_channels_TSP2011}
L.~Ruan and V.~Lau, ``Dynamic interference mitigation for generalized partially
  connected quasi-static {MIMO} interference channel,'' \emph{IEEE Transactions
  on Signal Processing}, vol.~59, no.~8, pp. 3788--3798, Aug. 2011.

\bibitem{Guillaud_Gesbert_Eusipco2011}
M.~Guillaud and D.~Gesbert, ``Interference alignment in the partially connected
  {K}-user {MIMO} interference channel,'' in \emph{Proc. European Signal
  Processing Conference (EUSIPCO)}, Barcelona, Spain, Aug. 2011.

\bibitem{Moallemi_MinSum_convergence_IT10}
C.~C. Moallemi and B.~V. Roy, ``Convergence of min-sum message-passing for
  convex optimization,'' \emph{IEEE Transactions on Information Theory},
  vol.~56, no.~4, pp. 2041--2050, Apr. 2010.

\end{thebibliography}

\end{document}